\documentclass[aps,pre,reprint,showpacs,superscriptaddress]{revtex4-1}

\usepackage[dvips]{graphicx}
\usepackage{dcolumn}
\usepackage{bm}
\usepackage{amsmath}
\usepackage{amsfonts}
\usepackage{amssymb}
\usepackage{natbib}
\usepackage[utf8]{inputenc}

\begin{document}


\title{Cyclones and attractive streaming generated by acoustical vortices}

\author{Antoine Riaud}
\affiliation{Institut d'Electronique, de Micro\'electronique et Nanotechnologie (IEMN), LIA LICS, Universit\'{e} Lille 1 and EC Lille, UMR CNRS 8520, 59652 Villeneuve d'Ascq, France}
\affiliation{CNRS UMR 7588, UPMC Universit\'{e} Paris 06, Institut des NanoSciences de Paris (INSP), F-75005, Paris, France}
\author{Michael Baudoin}
\affiliation{Institut d'Electronique, de Micro\'electronique et Nanotechnologie (IEMN), LIA LICS, Universit\'{e} Lille 1 and EC Lille, UMR CNRS 8520, 59652 Villeneuve d'Ascq, France}
\author{Jean-Louis Thomas}
\affiliation{CNRS UMR 7588, UPMC Universit\'{e} Paris 06, Institut des NanoSciences de Paris (INSP), F-75005, Paris, France}
\author{Olivier Bou Matar}
\affiliation{Institut d'Electronique, de Micro\'electronique et Nanotechnologie (IEMN), LIA LICS, Universit\'{e} Lille 1 and EC Lille, UMR CNRS 8520, 59652 Villeneuve d'Ascq, France}


\date{\today}

\begin{abstract}
Acoustical and optical vortices have attracted large interest due to their ability in capturing and manipulating particles with the use of the radiation pressure. Here we show that acoustical vortices can also induce axial vortical flow reminiscent of cyclones whose topology can be controlled by adjusting the properties of the acoustical beam. In confined geometry, the phase singularity enables generating "attractive streaming" with a flow directed toward the sound source.  This opens perspectives for contact-less vortical flow control. \end{abstract}


\maketitle


\section{Introduction \label{intro}}

Acoustic streaming, that is to say vortical flow generated by sound plays a fundamental role in a variety of industrial and medical applications such as sonochemical reactors \cite{iecr_kumar_2007},  megasonic cleaning processes \cite{pst_gale_2007}, ultrasonic processing \cite{jms_li_2004}, acoustophoresis \cite{loc_muller_2012}, or therapeutic ultrasound \cite{pt_baker_2001,jasa_crum_2005}. More recently, acoustic streaming is the subject of a burst of interest with the development of microfluidic applications \cite{loc_wiklund_2012,rmp_friend_2011}. For instance, it is at the core of the physics involved in  droplets actuation with Surface Acoustic Waves (SAW) \cite{arfm_yeo_2014} for lab-on-a-chip facilities, providing a versatile tool for droplet displacement \cite{abc_wixforth_2004,saa_renaudin_2006,pre_brunet_2010}, atomization \cite{pof_yeo_2008}, jetting \cite{jjap_shiokawa_1990,prl_tan_2009} or vibration \cite{pre_brunet_2010,apl_baudoin_2012,l_blamey_2013}. Moreover, vorticity associated with acoustic streaming is the main envisioned phenomenon to ensure efficient mixing of liquids \cite{apl_sritharan_2006,prl_frommelt_2008}.  

Different forms of streaming are generally distinguished according to the underlying physical mechanism \cite{asa_beyer_1997,arfm_ryley_2001}. \textit{Boundary layer-driven streaming} \cite{jasa_hamilton_2003} arises when an acoustic wave impinges a fluid/solid interface due to viscous stresses inside the viscous boundary layer. This form of streaming can be decomposed between \textit{inner streaming}, also called Schlichting streaming \cite{pz_schlichting_1932}, occurring inside the viscous boundary layer and counter rotating \textit{outer streaming} outside it \cite{ptrsl_rayleigh_1884}. The former is not exclusive to acoustics since it does not require compressibility of the fluid but only the relative vibration of a fluid and a solid. The latter, first enlightened by Lord Rayleigh, can be either seen as the fluid entrainment outside the boundary layer induced by Schlichting streaming or as a consequence of the tangential velocity continuity requirement for an acoustic wave at a fluid/solid boundary. Finally \textit{bulk streaming}, or so-called Eckart streaming \cite{Eckart} is due to the thermo-viscous dissipation of acoustic waves and the resulting pseudo-momentum transfer to the fluid \cite{prsla_piercy_1954, Sato_and_Fujii}. Since the early work of Rayleigh \cite{ptrsl_rayleigh_1884}, many studies have been dedicated to acoustic streaming and the investigation of the influence of various phenomena on the resulting flow, such as unsteady excitation \cite{az_rudenko_1971,ap_rudenko_1998,eif_sou_2011}, nonlinear acoustic wave propagation \cite{spa_romanenko_1960,spa_statnikov_1967} or high hydrodynamic Reynolds number \cite{jasa_menguy_1999,jasa_reyt_2013}. However, in all these studies, only plane of focalized acoustical waves \cite{ap_nyborg_1998} are considered.

In this paper, we report on bulk acoustic streaming generated by specific solutions of the Helmholtz equation called \textit{acoustical vortices}. New acoustic streaming configurations are obtained with cyclone-like flows, whose topology mainly depends on the one of the acoustical vortex. Flow streamlines are not only poloidal as in classic bulk streaming \cite{Eckart}, but also toroidal due to the orbital momentum transfer. This special feature provides an acoustical control of the axial vorticity, while in all forms of acoustic streaming reported up to now, the topology of the induced hydrodynamical vortices is mainly determined by the boundary conditions. Finally, in confined geometries, the azimuthal vorticity can also be tailored by adjusting the properties of the acoustic beam. In this way, \textit{attractor and  repeller hydrodynamic} vortices, corresponding respectively to flow directed toward or away from the sound source, can be obtained.

\section{Theoretical analysis}

Acoustical vortices (or Bessel beams) are helical waves possessing a pseudo orbital angular momentum and a phase singularity on their axis (for orders $\geq 1$). The pitch of the helix $l$ is called order or topological charge  \cite{ThomasBrunetCoulouvrat}. These waves are separated variables general solutions of the Helmholtz equation in cylindrical coordinates and are therefore not exclusive to acoustics (see e.g. \cite{Allen1992} for their optical counterparts). Separated variables solutions means that their axial and radial behavior are independant, i.e. the diffraction is canceled for infinite aperture and remains weak in others cases \cite{Durnin1987}. This enables their controlled synthesis even in confined geometries. Acoustical vortices can be generated by firing an array of piezo-electric transducers with a circular phase shift \cite{Hefner1999} or using inverse filtering techniques \cite{MarchianoThomas2003, prl_marchiano_2008, njp_brunet_2009}. As little as four transducers are enough to develop a first order vortex \cite{Hefner1999}. Recently, it has been observed that their orbital momentum can be transferred to dissipative  media which results in a measurable torque for solids \cite{He1995,Volke2008} or azimuthal rotation for fluids \cite{Anhaeuser}.  
\begin{figure}[htbp]
\includegraphics{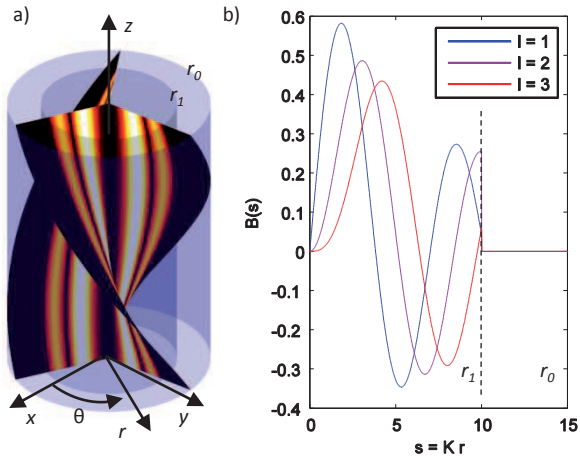}
\caption{\textbf{a:} Acoustical vortex with topological charge $l=3$, $\tan{(\alpha) = 1.21}$ and $K r_1 = 10$ ($z$-axis was dilated 10 times). Surfaces correspond to the phase $l\theta +k_z z = \pi/2$ while colors indicate the magnitude of the radial function $B$. \textbf{b:} Corresponding radial function $B(K r)$ for $l = 1$ to $3$. \label{fig: vortex}} 
\end{figure}

In the following, we derive the equations of the flow generated by an attenuated collimated \textit{Bessel beam} of finite radial extension $r_1$ (Fig. \ref{fig: vortex}.a), traveling along the $z$-axis of an unbounded cylindrical tube of radius $r_0$. This model constitutes an extension of Eckart's perturbation theory \cite{Eckart} initially limited to \textit{plane wave}. In the case of Bessel beams \cite{Hefner1999}, the density variation $\rho_1$ induced by the acoustical wave takes the form:
\begin{eqnarray}
\rho_1 (r,\theta,z,t) & = & \hat{\rho}_1 B(s) \sin(l \theta + k_z z - \omega t),
 \label{eq: rho1}\\
B(s) & = & A(s)J_l(s)
\label{eq: A and B}
\end{eqnarray}
In these equations, $\hat{\rho}_1$, $l$, $\theta$, $k_z$, $\omega$, $t$ and $J_l$ denote respectively the amplitude of the acoustical wave, the topological charge of the Bessel beam, the angular coordinate, the projection of the wave vector on the $z$-axis, the wave angular frequency, the time and the cylindrical Bessel function of order $l$. The spatial window function, $A(s)$, is used to limit the infinite lateral extension of Bessel function. The phase of such vortex is given by $\phi = l \theta +k_z z - \omega t$, yielding to helicoidal equiphase surfaces as shown in figure  \ref{fig: vortex}.a. We introduce the shorthand notation $s = K r$, and by analogy $ s_1 = K r_1 $, $s_0 = K r_0$, with $K$ the transversal component of the wave vector. It is defined by the dispersion relation of a Bessel beam:  $K^2 +k_z^2 = \omega^2/c^2$, with $c$ the sound speed. We also introduce the variable $\alpha$ measuring the helicoidal nature of the flow and defined by $\tan{(\alpha)}=k_z/K$. The radial dependence in equation (\ref{eq: rho1}) is based on Bessel functions, which are plotted in figure (\ref{fig: vortex}.b). Provided that $l \geqslant 1$, these functions cancel at $s = 0$, where destructive interference between the wavelets from opposite sides of the vortex occurs. Consequently, the core of the vortex is not solely a phase singularity, but also a shadow-area.

Following Eckart \cite{Eckart}, acoustic streaming can be calculated by decomposing the flow into a first order compressible and irrotational flow (corresponding to the propagating acoustical wave) and a second order incompressible vortical flow (describing the bulk acoustic streaming). The insertion of this decomposition into Navier-Stokes compressible equations yields Eckart's diffusion equation for the second order vorticity field $\vec{\varOmega}_2 = \vec{\nabla} \times \vec{u}_2$, with $\vec{u}_2$ the second order velocity field. This diffusion is forced by a nonlinear combination of first order terms and simplifies at steady-state into:
\begin{eqnarray}
\label{eq: Eckart}
& & \Delta \vec{\varOmega_2}  =  - \frac{b}{\rho_0^2} \vec{\nabla} \rho_1 \times \vec{\nabla} \frac{\partial \rho_1}{\partial t},
\end{eqnarray} 
with $b  =  4/3 + \mu'/\mu$, $\mu'$ the bulk viscosity, $\mu$ the shear viscosity, $\rho_0$ the density of the fluid at rest and $\rho_1$ the first order density variation. Since the streaming flow is incompressible, we can introduce the vector potential $\vec{\Psi}_2$ such that $\vec{u}_2 = \vec{\nabla} \times \vec{\Psi}_2$, with Coulomb gauge fixing condition: $\vec{\nabla} . \vec{\Psi}_2 = 0$. The resolution of equation (\ref{eq: Eckart}) thus amounts to the resolution of the inhomogeneous biharmonic equation:  $\Delta^2 \vec{\Psi}_2 = - \frac{b}{\rho_0^2} \vec{\nabla} \rho_1 \times \vec{\nabla} \frac{\partial \rho_1}{\partial t}$. Originally, this equation was integrated by Eckart for truncated plane waves. In the present work, we solve it in the case of Bessel beams, whose expression is given by equations (\ref{eq: rho1}) to (\ref{eq: A and B}). Owing to the linear nature of this partial differential equation, we consider only solutions verifying the symmetries imposed by the forcing term and the boundary conditions: no-slip condition on the walls, infinite cylinder in the z direction and no net flow along the channel. In this case, the problem reduces to a set of two linear ordinary differential equations, which were integrated with standard methods. The complete procedure is detailed in appendix.

Results are given by equations (\ref{eq: u_z}) to (\ref{eq: elastic energy}):
\begin{eqnarray}
u_2^z & = &  2 \frac{\varOmega^{\star}_\theta}{K} \left[  
\left(1 - \frac{s^2}{s_0^2} \right) f(s_0)  \right. \label{eq: u_z} \\
& + & \left. \frac{1}{2} \left(\frac{s^2}{{s_0}^2} \varLambda^l_z(s_0) -\varLambda^l_z(s) \right)  \right], \nonumber
 \\
u_2^{\theta} & = &  \frac{\varOmega^{\star}_z}{K}\left(\frac{s}{{s_0}^2} \varLambda^l_\theta(s_0) - \frac{1}{s}  \varLambda^l_\theta(s) \right), \label{eq: u_theta} 
\end{eqnarray}
\begin{eqnarray}
\mbox{with } \;  & & f(s)  =  - \frac{1}{2} \varLambda^l_z(s) + \frac{2}{s^2} \int_{0}^{s} x_1 \varLambda^l_z (x_1) dx_1,
\label{eq: definition of f}\\
& & \varLambda^l_\theta (s)  =  \int_{0}^{s} x_2\int_{0}^{x_2} \frac{B^2(x_1)}{x_1} dx_1 dx_2,
\label{eq: definition of Lambda theta} \\
& & \varLambda^l_z (s)  =  \int_{0}^{s} \frac{1}{x_2} \int_{0}^{x_2} x_1 B^2(x_1)dx_1 dx_2,
\label{eq: definition of Lambda z}\\
& & \varOmega^{\star}_{\theta}  =  \frac{1}{2}  \frac{\omega b \tan{(\alpha)}}{\rho_0 c^2}E_1,
\label{eq: K_theta}\\
& & \varOmega^{\star}_z  =  \frac{1}{2} \frac{\omega b l}{\rho_0 c^2} E_1,
\label{eq: K_z}\\
& & E_1  =  c^2 \frac{(\hat{\rho}_1)^2}{\rho_0}.
\label{eq: elastic energy}
\end{eqnarray}
 In these expressions, we see that the ratio between the axial and azimuthal velocities $u_2^z/u_2^\theta$ is proportional to the ratio $\varOmega^{\star}_\theta/\varOmega^{\star}_z = \tan{(\alpha)}/l$, indicating that as $\alpha$ decays or $l$ grows up (increasing the gradients along $r$ and $\theta$ directions respectively), the azimuthal velocity tends to dominate over its axial counterpart.  Both speeds are proportional to the acoustic energy rather than the amplitude, emphasizing the fundamental nonlinear nature of acoustic streaming. Furthermore, both terms are linearly proportional to $\omega$ such that its product with the elastic potential energy (\ref{eq: elastic energy}) refers to the power flux carried by the wave.     

\begin{figure}[htbp]
\includegraphics{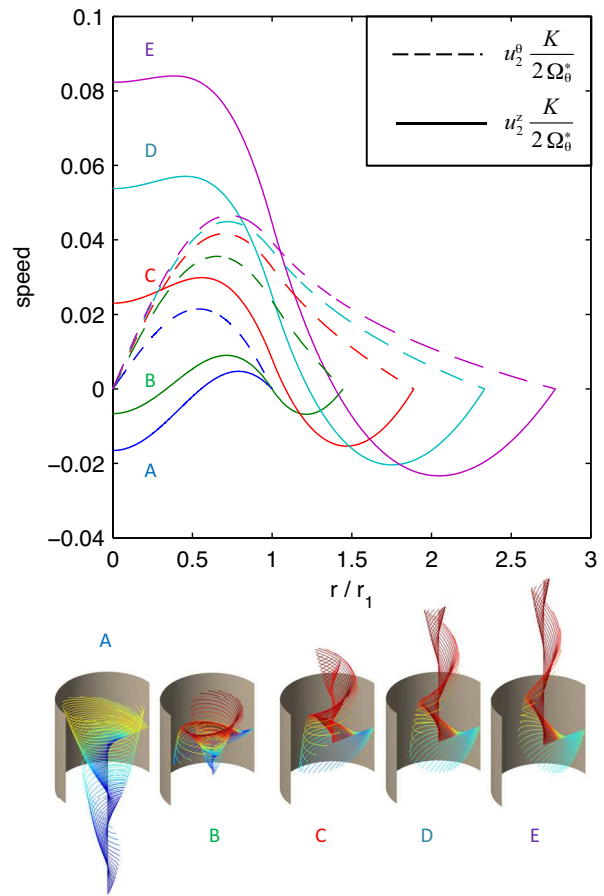}
\caption{\textbf{Top:} Non-dimensional velocities for $l=1$, $\tan{(\alpha) = 1.21}$ and $K r_1 = 1.84$ for progressively increased cavity geometrical proportions $r_0/r_1 = $ [1(A)\quad 1.44(B)\quad  1.89(C)\quad  2.33(D)\quad 2.78(E)]. Axial velocity is represented by solid lines and the azimuthal component by the dashed ones. \textbf{Bottom:} Flow streamlines.  Colors are indicative of the speed magnitude along $u_2^z$: extrema are represented by the most intense colors, red for positive and blue for negative. \label{fig: changing r_0/r_1}} 
\end{figure}

Equations (\ref{eq: u_z}) to (\ref{eq: elastic energy}) were integrated numerically to compute the velocity field. A square spatial window function for $A(s)$ (whose expression is given in appendix \ref{A_Eck}) is chosen to simplify the algebra. In the following, we investigate the case $l=1$, $\tan{(\alpha) = 1.21}$ and $K r_1 = 1.84$ to get an overview of the flow pattern when the geometric ratio $r_0/r_1$ is tuned. Resulting velocity profiles and the associated streamlines are presented in figure (\ref{fig: changing r_0/r_1}). They show a combination of axial and azimuthal vortical structures whose topology depends on the ratio $r_0/r_1$.

\section{Repeller and attractor vortices}

It is commonly accepted that Eckart's streaming is the result of pseudo-momentum transfer from the sound wave to the fluid \cite{prsla_piercy_1954}. Consequently, the acoustic beam ($r<r_1$) should push the fluid away from the transducer. This is what actually occurs in weakly confined geometry, that is to say for the largest ratios $r_0/r_1$ (see Fig. \ref{fig: changing r_0/r_1} C to E). In these cases, confinement and mass conservation impose a back-flow at the periphery of the acoustic beam, resulting in azimuthal vorticity similar to the one observed by Eckart. But Bessel beams also carry an angular momentum, which is transmitted to the fluid and results in axial vorticity \cite{Anhaeuser}. Since for $l>0$ the wave is rotating in the positive direction (when time increases, equiphase is obtained for growing $\theta$), the azimuthal velocity is also positive. 

However, this analysis doesn't hold when applied to very confined geometries such as A and B, where the beam covers almost all the cylindrical channel. Under these conditions, radial variations of the beam intensity must be considered. Indeed, in figure (\ref{fig: vortex}) we clearly see that the Bessel beam offers a shadow-area in the neighborhood of its axis, where the wave amplitude cancels. This holds for all non-zero orders vortices. The backflow generally appears where the wave forcing is weaker. Hence, the fluid recirculation can either occur near the walls or at the core of the beam, which becomes the only option as the free-space at the periphery of the vortex shrinks to 0, as in case A. Let's call these vortices \textit{attractor vortices} since they tend to drive fluid particles towards the sound source, and their opposite \textit{repeller vortices}, since they push fluid particles away from the source. Although streaming pushing the fluid away from a transducer is common, (i) it is not usually associated with axial vorticity and (ii) the vorticity topology depends on the boundary conditions. Furthermore Bessel beams enable for the first time the synthesis of attractive vortices, offering original prospects for flow control and particle sorting in confined geometries.

\begin{figure}[htbp]
\includegraphics{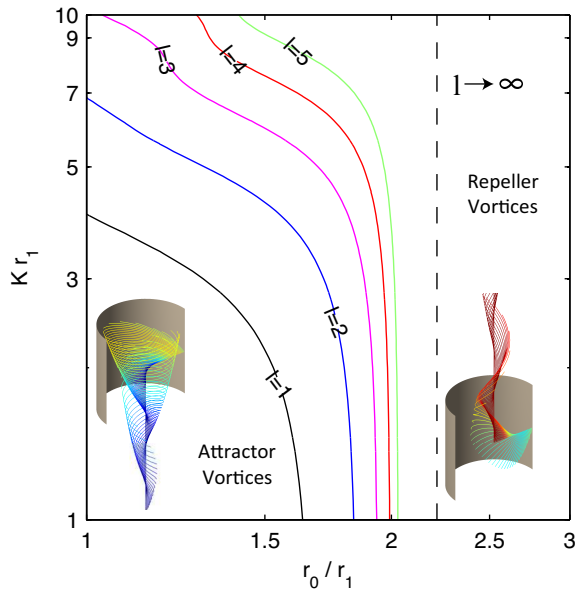}
\caption{Contour plot of $u^z_2(r=0)=0$ at various topological charge $l$,  typical dimension $K r_1$ and geometrical ratio $r_0/r_1$. Parameters plane is partitioned into two areas, one close to the origin corresponding to \textit{attractor} vortices with negative axial velocities at $r=0$ and the other one corresponding to \textit{repeller} vortices. The dashed line at $r_0/r_1 \sim 2.218$ indicates an asymptotic limit obtained for large $l$ values. \label{fig: vortex_map}} 
\end{figure}

Intrigued by this reverse-flow motion, we performed a systematic investigation on the conditions of its appearance. Looking at the expression of the velocity, we notice that the sign of $u_2^z(r=0)$ is independent of $\tan(\alpha)$, such that the set of parameters reduces to the topological charge $l$,  the typical dimension $K r_1$ and the geometrical ratio $r_0/r_1$. All these parameters were gathered in figure (\ref{fig: vortex_map}) to give an overview of the streaming induced by Bessel beams in confined space.   
Looking at the flow map for $l=1$, we first notice that there is a bounded set of parameters leading to attractor vortices. Indeed, these vortices are squeezed by two restrictions: the beam must be confined enough (ratio $r_0/r_1$ close to one) as previously explained, and the value of $K r_1$ has to be small. Looking back at figure (\ref{fig: vortex}.b), we notice that as $K r_1$ increases, the Bessel function amplitude decreases on the periphery which facilitates the flow recirculation close to the channel walls. This trend is reinforced by the apparition of new nodes of the Bessel function for higher values of $K r_1$ and the quadratic dependence of the streaming flow. In addition, as the beam gets wider, the envelope of the beam weakens for increasing $r$ and hence, the recirculation preferentially flows towards the periphery.

Introducing the topological order $l$ as a free parameter, we notice the progressive broadening of the attractor domain. Referring to figure (\ref{fig: vortex}.b), it appears that Bessel functions of higher order roughly translate towards increased $K r_1$, or reciprocally need a higher $K r_1$ to reach the analog extremum. This explains the $K r_1$ part of the broadening, whereas the $r_0/r_1$ is due to the progressive flattening of Bessel functions, which nonetheless rapidly saturates. Using the asymptotic forms of Bessel development, we computed this limit in the appendix. The extreme value is given solving the equation $\ln(x) = 1-1/x^2, \quad \mbox{with} \quad x = r_0/r_1$. The existence of this upper bound highlights the essential condition of the confined nature of the channel.

To compute these last results, we use a window function, $A(s)$, with a sharp cut-off to ease the comparison with Eckart results. If we relax this condition, no change is expected in the case of weakly confined beams, $Kr_1 >>1$. For such beam, the flow will recirculate preferentially at the periphery due to the radial decreasing of the Bessel function. The strictly confined case $r_0/r_1=1$ is possible since Bessel beams are the modes of cylindrical wave guides for discrete values of the radial wave number $K=s/r$, i.e no window A(s) is required. Hence flow reversal at the vortex core should be observable. The intermediate situation of strongly confined beam, $1 < r_0/r_1=1 <2$, is more challenging to carry out experimentally due to diffraction spreading. However, this problem is mitigated since truncated Bessel beam are weakly diffracting \cite{Durnin1987}.



\section{Conclusion}

In this paper, we derive the streaming flow induced by Bessel beams (acoustical vortices). The resulting flow topology is reminiscent of cyclones with both axial and azimuthal vorticity. The axial component is solely controlled by the acoustic field. Regarding the azimuthal vorticity, two categories of flow pattern should be distinguished: \textit{repeller and attractor vortices}. The first category exhibits a positive velocity at the center of the beam, and appears when the beam radius is small compared to the fluid cavity; whereas the latter needs a very confined geometry, and develops negative velocity in its core. To the best of our knowledge, streaming-based \textit{attractor beams} have never been described before and are due to the specific radial dependence of the sound wave intensity in Bessel beams. This work opens prospects for vorticity control, which is an essential feature in many fluidic systems \cite{ctam_ottino_1989, s_chorin_1994, pof_gmelin_2001,jfm_zhu_2002}. Moreover, the combination of attractive streaming and radiation pressure \cite{jasa_baresh_2013,settnes_pre_2012,pre_zhang_2011} induced by acoustical vortices could provide an efficient method for particles sorting. Indeed, large particles are known to be more sensitive to radiation pressure and small particles to the streaming \cite{loc_hagsater_2007}. Large particles would therefore be pushed away from the sound source by the radiation pressure while small particles would be attracted by the flow toward it. Compared to existing techniques relying on radiation pressure generated by standing waves \cite{loc_petersson_2005,apl_wood_2008}, the advantage would be that a resonant cavity is not mandatory to sort particles with acoustical vortices since progressive waves can be used. 

\acknowledgments
This work was supported by ANR project ANR-12-BS09-0021-01.

\appendix

\section{Resolution of Eckart equation for acoustical vortices}
Eckart acoustic streaming \cite{Eckart} is adequately described by a set of non-linear partial differential equations. Although exact analytical solutions have not been found in the general case, the problem can be solved with a perturbation analysis, as long as the acoustic wave propagation is weakly nonlinear (weak acoustical Mach Number) and the flow remains laminar (weak Reynolds number). Following Eckart, the flow generated by a transducer can be decomposed into a first order compressible and irrotational flow (corresponding to the propagating acoustic wave) and a second order incompressible  vortical flow (corresponding to the acoustic streaming) \footnote{NB: In his seminal paper, Eckart also computed the irrotational part of the second order flow corresponding to the effect of nonlinearities on the acoustic wave propagation. Then, using Helmoltz decomposition, he focused on the incompressible part, corresponding to the acoustic streaming. Here we do not consider the effect of nonlinearities on the acoustic wave propagation.}:
\begin{eqnarray}
\rho & = & \rho_0 + \rho_1+\rho_2+...\\
\vec{u} & = & \vec{u}_1+\vec{u}_2+...
\end{eqnarray}
with $\rho_2 \ll \rho_1 \ll \rho_0$ and $\| \vec{u_2} \| \ll \| \vec{u_1} \|$. Basically the order of magnitude of the ratio between first order and second order fields is given by the acoustical Mach number. In this development, we have considered a homogeneous fluid at rest in the absence of the acoustic field. Thus the density $\rho_0$ is constant in space and time, and the velocity $\vec{u_0} = \vec{0}$. 

By replacing this decomposition into Navier-Stokes compressible equations, Eckart showed that the first order field is solution of D'Alembert (wave) equation. Acoustical vortices  are solution of this equation in cylindrical coordinates \cite{McGloin} and their expression calculated by Hefner and Marston \cite{Hefner1999} takes the following form for weakly attenuated waves :
\begin{equation}
\rho_1 (r,\theta,z,t) = \hat{\rho}_1 A(K r) J_l(K r) \sin(l \theta + k_z z - \omega t)
 \label{rho1}
\end{equation}
In this equation, $\phi = l \theta +k_z z - \omega t$ is the phase of the acoustical vortex, $l$ the topological charge of the vortex, $\theta$ the angular coordinate, $k_z$ the projection of the wave vector on $z$-axis, $z$ the height, $\omega$ the wave frequency, $t$ the time. Finally, $\hat{\rho_1}$ is the amplitude of the first order density fluctuation, which is related to its pressure counterpart $\hat{P}_1$ according to $\hat{\rho_1} = \hat{P_1}/c^2$ and $K$ the transversal component of the wave vector. It is defined by the dispersion relation of an acoustical vortex:  $K^2 +k_z^2 = \omega^2/c^2$, with $c$ the sound speed. 

Eckart obtained in his paper a diffusion equation for the second order vorticity field $\vec{\varOmega}_2 =  \vec{\nabla}  \times \vec{u}_2$, which can be used to compute the acoustic streaming. In the following, we consider steady streaming generated by a monochromatic acoustic wave with constant amplitude  and therefore Eckart equations reduces to:
\begin{eqnarray}
\Delta \vec{\varOmega_2} & = & - \frac{b}{\rho_0^2} \vec{\nabla} \rho_1 \times \vec{\nabla} \frac{\partial \rho_1}{\partial t}  \label{Eckart}\\
 b & = & 4/3 + \mu'/\mu
\end{eqnarray}
with $\mu$ the shear viscosity and $\mu'$ the bulk viscosity.
From now on, we will use the shorthand notation $s = K r$, and $s_1 = K r_1$, $s_0 = K r_0$. Besides, we introduce $B$ to gather the radial dependence of the beam:
\begin{equation}
B(s)  =  A(s)J_l(s) \\
\label{A and B}
\end{equation}
where the function, $A(s)$, is introduced to limit the infinite lateral extension of Bessel function.The derivation of $\rho_1$ in equation (\ref{rho1}) in cylindrical coordinates, and the replacement of the result into equation (\ref{Eckart}) gives a inhomogeneous Poisson equation with the first order field playing the role of the streaming source term:
\begin{eqnarray}
& & \frac{1}{K^2} \Delta \vec{\varOmega}_2(r,\theta,z)  =  \varOmega_{\theta}^\star \frac{d B^2(s)}{d s} \vec{e_{\theta}} -  \frac{\varOmega_z^\star}{s} \frac{d B^2(s)}{d s} \vec{e_z}  \label{Eckart vortex}\\
& & \varOmega_{\theta}^\star  =  \frac{1}{2}  \frac{k_z \omega b}{K \rho_0 c^2}E_1\\
& & \varOmega_z^\star  =  \frac{1}{2} \frac{\omega b l}{\rho_0 c^2} E_1\\
& & E_1  =  c^2 \frac{{\hat{\rho}_1}^2}{\rho_0}
\label{elastic energy}
\end{eqnarray}
The beam is assumed to be of infinite extent along $z$ and invariant by rotation $\theta$ around this axis, therefore $\vec{\varOmega}_2$ has only a radial dependence. Besides, the conservative nature of vorticity allows us to drop-off the $\vec{e_r}$ component. The resulting solution candidate for  $\vec{\varOmega}_2$ is:
\begin{equation}
 \vec{\varOmega}_2 = \varOmega_2^\theta(s) \vec{e_{\theta}} +  \varOmega_2^z(s) \vec{e_{z}}
 \label{proposed solution} 
\end{equation}
Plugging it into equation (\ref{Eckart vortex}) gives two linear ODEs:
\begin{eqnarray}
& & s^2\frac{d^2}{d s^2}\varOmega_2^\theta + s \frac{d}{d s}\varOmega_2^\theta  - \varOmega_2^\theta  =  s^2 \varOmega_{\theta}^\star \frac{d B^2(s)}{d s}\\
& & s\frac{d^2 }{d s^2} \varOmega_2^z + \frac{d }{d s}\varOmega_2^z  =  - \varOmega_z^\star \frac{d B^2(s)}{d s}
\label{ODE Eckart streaming}
\end{eqnarray}
Using standard methods, the homogeneous ($H$) and particular ($P$) solutions are determined:
\begin{eqnarray}
& & {\varOmega_2^\theta}|_{H} = M_1^{\theta} s + \frac{N_1^\theta}{s}
\label{solution homogene Omega2theta} \\
& & {\varOmega_2^\theta}|_{P} = \frac{1}{s} \varOmega_{\theta}^\star  \int_{0}^{s} x_1 B^2(x_1)dx_1
\label{solution particuliere Omega2theta} \\
\end{eqnarray}
The equation along $z$ is treated by introducing $g = \frac{d}{d s} \varOmega_2^z$
\begin{eqnarray}
g|_H & = & \frac{M_{z}^1}{s} \\
g|_P & = & -\varOmega_z^\star \frac{B^2(s)}{s} \\
\varOmega_2^z & = & N_1^z + M_{z}^1 \ln(s) - \varOmega_z^\star  \int_{0}^{s} \frac{B^2(x_1)}{x_1}dx_1
\label{solution Omega 2z}
\end{eqnarray}
Removing the terms diverging at $s=0$, we have:
\begin{eqnarray}
 \vec{\varOmega}_2  & = &  \left[ M_1^{\theta} s +  \frac{1}{s} \varOmega_{\theta}^\star \int_{0}^{s} x_1 B^2(x_1)dx_1 \right]\vec{e_{\theta}} \\
    & & + \left[ N_1^z - \varOmega_{z}^\star  \int_{0}^{s} \frac{B^2(x_1)}{x_1}dx_1 \right] \vec{e_{z}}  
 \label{Omega 2 solution}
\end{eqnarray}
Since the second order flow (streaming) is incompressible, we can introduce the vector potential $\vec{\Psi}_2$ verifying $\vec{u}_2 = \vec{\nabla} \times \vec{\Psi}_2$ with the gauge $\vec{\nabla} . \vec{\Psi}_2 =  0$ to compute the velocity field from the vorticity field:
\begin{equation}
\Delta \vec{\Psi}_2 = - \vec{\varOmega_2}
\label{Irrotational flow}
\end{equation}
For symmetry reasons, the flow is assumed to be invariant by rotation $\theta$ around $z$ and translation along the propagation axis $z$, and due to the conservative nature of $\vec{u}_2$, the radial component is dropped off. Consequently, the velocity field is of the form: $\vec{u}_2 = u_2^\theta(s)\vec{e_{\theta}} +u_2^z(s)\vec{e_z}$. Computing the curl of $\vec{\Psi}$ in order to get $\vec{u}_2$, we notice that $\vec{\Psi}_2 = \Psi_{\theta}(s)\vec{e_{\theta}} + \Psi_{z}(s)\vec{e_z}$. Equation (\ref{Irrotational flow}) is very similar to (\ref{Eckart}), except the source term: 
\begin{eqnarray}
& & s^2 \Psi_{\theta}'' + s \Psi_{\theta}'  - \Psi_{\theta}  =  - \frac{1}{K^2} \left(  M_1^{\theta} s^3 + s\varOmega_{\theta}^\star \int_{0}^{s} x_1 B^2(x_1)dx_1 \right) \nonumber \\
& & s \Psi_{z}'' + \Psi_{z}'  =  \frac{s}{K^2} \left( - N_1^z   + \varOmega_{z}^\star \int_{0}^{s} \frac{B^2(x_1)}{x_1}dx_1 \right) \nonumber
\end{eqnarray}
Using the same procedure as for $\vec{\varOmega}_2$ we get the general solution:
\begin{eqnarray}
\Psi_{\theta} & = & M_{\theta}^2 s - \frac{1}{K^2} \left( \frac{M_1^{\theta}}{8} s^3 + \frac{\varOmega_{\theta}^\star I_\theta(s)}{s} \right)\\
\Psi_{z} & = & N_z^2 + \frac{1}{K^2} \left( - N_1^z \frac{s^2}{4} + \varOmega_{z}^\star  I_z(s) \right) \\
I_\theta & = & \int_{0}^{s} x_3 \int_{0}^{x_3} \frac{1}{x_2} \int_{0}^{x_2} x_1 B^2(x_1)dx_1 dx_2 dx_3 \\
I_z & = & \int_{0}^{s} \frac{1}{x_3} \int_{0}^{x_3} x_2\int_{0}^{x_2} \frac{B^2(x_1)}{x_1} dx_1 dx_2 dx_3
\end{eqnarray}
The resulting velocity field can now be simply obtained by taking the curl of~$\vec{\Psi_2}$:
\begin{eqnarray}
u_2^\theta & = &  \frac{1}{K}\left( \frac{1}{2} N_1^z s - \frac{1}{s} \varOmega_{z}^\star \varLambda^l_\theta(s) \right) \\
u_{z}^{(2)} & = &  2 M_{\theta}^2 K - \frac{1}{K} \left( \frac{M_1^{\theta}}{2} s^2 +  \varOmega_{\theta}^\star \varLambda^l_z(s) \right) \\
\varLambda^l_\theta (s) & = & \int_{0}^{s} x_2\int_{0}^{x_2} \frac{B^2(x_1)}{x_1} dx_1 dx_2 \label{definition of Lambda theta} \\
\varLambda^l_z (s) & = & \int_{0}^{s} \frac{1}{x_2} \int_{0}^{x_2} x_1 B^2(x_1)dx_1 dx_2
\label{definition of Lambda z}
\end{eqnarray}
This velocity field must satisfy the adherence boundary condition at the wall of the channel $s = s_0$:
\begin{eqnarray}
u_2^\theta(s_0) & = & 0 \\
u_2^z(s_0) & = & 0 
\label{adherance and flow BC}
\end{eqnarray}
Besides, the steady and incompressible nature of the flow must not violate mass conservation, such that a closure condition is enforced:
\begin{equation}
\renewcommand\arraystretch{2}
\begin{array}{llcr}
\quad & \int_{0}^{2 \pi} \int_{0}^{r_0} \rho_0 u_2^z(r) r dr d \theta & = & 0 \\
\Leftrightarrow & \int_{0}^{s_0} x_1 u_z(x_1) dx_1 & = & 0
\end{array}
\end{equation}
The determinant of the system is equal to $\frac{s_0^4}{8K}$, such that it always admits a unique solution. Solving this linear system of equations, we get:
\begin{eqnarray}
N_1^z & = & \frac{2 \varOmega_z^\star}{{s_0}^2} \varLambda^l_\theta(s_0) \\
M_\theta^1 & = & \frac{4 \varOmega_\theta^\star}{{s_0}^2} \left(f(s_0)- \frac{1}{2}\varLambda^l_z(s_0) \right)   \\
M_\theta^2 & = & \frac{\varOmega_\theta^\star}{K^2} f(s_0)\\
f(s) & = & - \frac{1}{2} \varLambda^l_z(s) + \frac{2}{s^2} \int_{0}^{s} x_1 \varLambda^l_z (x_1) dx_1 
\label{definition of f}
\end{eqnarray}
Including these boundary conditions in the expressions of the velocity field, we finally obtain:
\begin{eqnarray}
& & u_2^\theta  =   \frac{\varOmega_z^\star}{K}\left(\frac{s}{{s_0}^2} \varLambda^l_\theta(s_0) - \frac{1}{s}  \varLambda^l_\theta(s) \right) \nonumber \\
& & u_{z}^{(2)}  =   2 \frac{\varOmega_\theta^\star}{K} \left(\left(1 - \frac{s^2}{{s_0}^2} \right) f(s_0)  + \frac{1}{2} \left(\frac{s^2}{{s_0}^2} \varLambda^l_z(s_0) -\varLambda^l_z(s) \right) \right) 
\nonumber
\end{eqnarray}

\section{Asymptotic development when $ K r_0 << 2 \sqrt{l+1}$}

In this section, we compute an asymptotic development of our final expression when  $ K r_0 << 2 \sqrt{l+1}$. We show that Eckart's result obtained for plane wave can be recovered as an asymptotic limit of our more general expression. Recovering, the Eckart result dictates the choice of the function $A(s)$:
\begin{equation}
A(s)  =  \begin{cases} 1,  \mbox{if } s < s_1 \\ 0,  \mbox{if } s \geqslant s_1 \end{cases}
\label{A_Eck}
\end{equation}

\subsection{Asymptotic development}

For all $s < s_0 = K r_0$, we have:

\begin{eqnarray}
& & J_l(s)  \sim  \frac{1}{l!}\,{\left( \frac{s}{2}\right) }^{l}\\
& & \Lambda^l_z(s)  \sim  \begin{cases} \frac{{s}^{2\,l+2}}{{\left( 2\,l+2\right) }^{2}\,{2}^{2\,l}\,{l!}^{2}}, & \mbox{if } s < s_1 \\ {s_1}^{2\,l+2}\, \frac{1+ (2 l + 2 )\mathrm{ln}\left( s\right/s_1) }{\left( 2\,l+2\right) ^2 \,{2}^{2\,l}\,{l!}^{2}}, & \mbox{if } s \geqslant s_1 \end{cases} \\
%
%
& & f  \sim  \begin{cases} \frac{{s}^{2\,l+2}}{{\left( 2\,l+2\right) }^{2}\,\left( 2\,l+4\right) \,{2}^{2\,l}\,{l!}^{2}}(2-\frac{2l+4}{2}), & \mbox{if } s < s_1 \\ \\
 C_l \left[ \begin{array}{l}
        (s_1/s)^2 \left( l\,(1-(s/s_1)^{2}\,)+ \frac{1}{l+2} \right) \\
        + \left( (l + 1 )\mathrm{ln}\left( s\right/s_1) -(1/2) \right)
    \end{array}\right] , & \mbox{if } s \geqslant s_1 
\end{cases}\\
& &\mbox{with} C_l = \frac{{s_1}^{2\,l+2}\,}{\left( 2\,l+2\right) ^2 \,{2}^{2\,l}\,{l!}^{2}}
\end{eqnarray}

\subsection{Recovering Eckart's streaming with $l=0$ and $ K r_0 << 1$}

The case of plane wave can be recovered from our expression by considering a topological charge equal to zero and a radius $r_0 \ll 1/K$:

\begin{eqnarray}
& & \Lambda^l_z(s)  \sim  
\begin{cases} s^2/4, & \mbox{if } s < s_1 
\\ (s_1^2/4) (1+ 2 \mathrm{ln}\left( s\right/s_1)), & \mbox{if } s \geqslant s_1 
\end{cases} \\
& & f  \sim  \begin{cases} 0, & \mbox{if } s < s_1 
\\  \\(s_1^2/8) \left[(s_1/s)^2+ 2 \mathrm{ln}\left( s\right/s_1) -1 \right], & \mbox{if } s \geqslant s_1 
\end{cases} 
\end{eqnarray}
In the original paper \cite{Eckart}, Eckart introduces the notation $x = s/s_0$ and $y = s_1/s_0$
\begin{equation}
u_2^z \sim 2 \frac{\varOmega^\star_\theta}{K} \begin{cases}
s_1^2/4\left[ (1/2)(1-(x/y)^2) - (1-y^2/2)(1-x^2) - \ln(y) \right], & \mbox{if } s < s_1 \\
\\
- s_1^2/4\left[ (1-y^2/2)(1-x^2)+\ln(x)  \right], & \mbox{if } s  \geqslant s_1
\end{cases} \label{Eckart_l0}
\end{equation}
Equation (\ref{Eckart_l0}) is exactly the expression of the acoustic streaming obtained by Eckart \cite{Eckart} for plane waves.

\subsection{Asymptotic limit for large values of $l$ and $ K r_0 << 2 \sqrt{l+1}$}

\begin{eqnarray}
& & \Lambda^l_z(s)  \sim  \begin{cases} \frac{{s}^{2\,l+2}}{{\left( 2\,l\right) }^{2}\,{2}^{2\,l}\,{l!}^{2}}, & \mbox{if } s < s_1 \\ {s_1}^{2\,l+2}\, \frac{\mathrm{ln}\left( s\right/s_1) }{\left( 2\,l\right) \,{2}^{2\,l}\,{l!}^{2}}, & \mbox{if } s \geqslant s_1 \end{cases} \\
%
%
& & f  \sim  \begin{cases} -\frac{{s}^{2\,l+2}}{2 {\left( 2\,l\right) }^{2}\, {2}^{2\,l}\,{l!}^{2}}, & \mbox{if } s < s_1 \\ 
C_l \left[ \begin{array}{l}
        (s_1/s)^2(1-(s/s_1)^{2}) \\
        +  \mathrm{ln}\left( s\right/s_1)
    \end{array}\right] , & \mbox{if } s \geqslant s_1 
\end{cases}\\
& & C_l  =  \frac{{s_1}^{2\,l+2}\,}{4 l \,{2}^{2\,l}\,{l!}^{2}}\\
& & u_2^z(s=0)  =  2 \frac{\varOmega^\star_\theta}{K}  f(s_0) \\
\end{eqnarray}
Using Eckart \cite{Eckart} notation $y = s_1/s_0$:
\begin{equation}
u_2^z(s=0) = 2 \frac{\varOmega^\star_\theta C_l }{K} \left( y^2 -1   - \ln(y)\right)
\label{asymptotic velocity}
\end{equation}
We highlight here that in equation (\ref{asymptotic velocity}) $C_l$ is decreasing extremely fast such that increasing $l$ dramatically decreases the magnitude of $u_2^z(0)$.


%

\end{document}